# Agents All the Way Down

## A Methodology for Building Custom AI Agents from Substrate to Production

Marc Alier Forment[1], Juanan Pereira[2], Francisco José García-Peñalvo[3], María José Casañ Guerrero[1]

[1] Universitat Politècnica de Catalunya (UPC), Barcelona, Spain [2] Universidad del País Vasco / Euskal Herriko Unibertsitatea (UPV/EHU), Donostia-San Sebastián, Spain [3] Universidad de Salamanca (USAL), Salamanca, Spain

marc.alier@upc.edu

---

## Abstract

Thousands of software engineers are building or about to build **custom AI agents** — agents that live inside their own application, talk to their own data and tools, enforce their own security boundaries, and carry their own brand and audit trail. What separates custom agents from the general-purpose tier (Claude Code, Cursor, and the like) is *fit*, not capability: each is built for one job, by the engineer who will maintain it. No published practice, though, sets out how to build one end to end. The pieces are everywhere: function-calling APIs, the Model Context Protocol, general-purpose code agents to pair with. The practice that chains them together is not — it lives in podcasts, blog posts, and leaked system prompts. This paper writes that practice down as a methodology, **Agents All the Way Down**, with a defining shape: two preconditions the engineer crosses once and keeps, then three practices the engineer repeats for the life of the agent. The preconditions are **(P1) Substrate** — treating the large language model (LLM) as a software component, framed as `tools → system → messages` under today's prompt-caching schemes — and **(P2) Building blocks** — function calling, the Model Context Protocol (MCP), command-line interface (CLI) orchestration, the *liteshell* pattern, the agent loop, skills, characters, hooks, and scaffolding. The three repeated practices are **(P3)** prototype with a general-purpose agent; **(P4)** harvest, fold, and ship the result as a CLI — the *Turtle pattern*; and **(P5)** agent-tests-agent, in which a general-purpose agent drives the custom agent through behavioural scenarios. P5 complements classical software testing; it does not replace it. Once the preconditions are in hand, the engineer's working life is the loop P3 → P4 → P5 → P3 — and that loop is where the title lives. One corollary falls out for free: multi-agent orchestration is just CLI composition. We build the methodology framework-free, for engineering reasons rather than pedagogical ones (§7.1). It was distilled from the AAC (Agent-Assisted Creator), a custom agent for the open-source LAMB (Learning Assistants Manager and Builder) educational platform: built in about ten days by one developer with an AI pair-programmer, in production at two universities (~200 educator-creators) since April 2026, and since carried to other LAMB subprojects. We present it as a transferable practice, independent of any language or framework.

---

# 1. Introduction

### 1.1 What is a custom AI agent?

A **custom AI agent** is a piece of software that decides, on each turn, which tool to call next, with the help of a large language model. It is not a chatbot or a generic copilot: it is purpose-built — for one application, one domain, one set of users — by the engineer who will maintain it.

The general-purpose code-agent tier — Claude Code [Anthropic 2026a], OpenCode [Anomaly 2026], Cursor [Cursor 2026], Aider [Gauthier 2026] — has become extraordinary engineering. Those are the **carpenter's bench saw**: powerful, broadly useful, deliberately general. The custom agent is the **cabinet maker's jig**: built for one specific job, by/for the cabinet maker, fitted to one piece of work. The distinction this paper concerns itself with is fit, not capability.

What makes an agent custom? Six axes:

1. **Domain knowledge embedded in the system prompt and skills** — the agent knows the engineer's codebase, conventions, vocabulary, and data shape as durable cached system content, not as conversational re-priming each session.
2. **Tools the general agent does not have** — internal APIs, private CLIs, line-of-business data sources. The agent operates on the engineer's *application's* surface rather than the host filesystem and shell.
3. **Deployment inside the engineer's own application** — the agent runs inside the FastAPI route, the Spring Boot service, the educational platform, the e-commerce backend. The user never sees a terminal; they see the product, with an agent in it.
4. **Security boundaries the engineer controls** — an allow-list fitted to the job; the agent cannot do what it does not need to do.
5. **Cost predictability through right-sized model choice** — small specialised model for the classifier, frontier model for the planner, local model for high-volume mechanical work. The cost curve is the engineer's to shape.
6. **Brand, voice, audit trail, and compliance posture** — the agent talks to *the engineer's* users in *the engineer's* voice; it logs to *the engineer's* audit pipeline; it complies with *the engineer's* regulatory boundaries.

An agent that needs at least one of those is a candidate for the custom build. An agent that needs none of them should not be built — the general-purpose tier handles it.

### 1.2 Why build one?

The engineer builds a custom agent when one or more of the six axes prevents the general-purpose agent from doing the job cleanly. Four concrete shapes of problem recur:

- **An assistant inside an educational platform** — the AAC (Agent-Assisted Creator) studied as a worked example in §6. Educators are not engineers; the agent must speak the language of pedagogy, not of code; it must operate the platform's surface, not the host's shell; it must run inside the platform's backend, not in a terminal.
- **A continuous-integration / continuous-deployment (CI/CD) helper** that knows the team's deployment topology, runbook conventions, and incident history. The general agent knows generic CI; the custom agent knows the team's.
- **A customer-service agent** that operates on the customer's own state. The general agent cannot resolve the customer record; the custom agent's first tool does.
- **A data-extraction pipeline agent** that handles documents of a shape the business sees a thousand times a day. The general agent can do one document well; the custom agent does the thousand reliably, at the right cost, inside the engineer's product.

The general-purpose agent can do any of these once; the custom agent does them reliably, at predictable cost, inside the engineer's application. **A custom agent is what the engineer builds when the general agent ends and the problem keeps going.**

### 1.3 How does the engineer build one?

The question, then, is: *what is the practice for building such a thing?* In 2026 the toolbox is extraordinary — function-calling APIs across every major LLM provider, the Model Context Protocol [Anthropic 2024b; MCP 2025] for tool exposure, mature general-purpose code agents usable as pair-programmers, dozens of agent frameworks. But the *practice* of building a custom agent has remained largely tacit. It lives in podcast interviews with framework creators, in leaked codebases — such as the Claude Code system-prompt leak of April 2026 — in GitHub READMEs that show what to do but not why, and in conference talks that present finished products without the path that produced them. **The ingredients have been catalogued; the recipe has not.**

This paper proposes one we have tested. The methodology has *five phases*, with an organising asymmetry that shapes the rest of the paper: **the first two phases are preconditions the engineer crosses once and keeps; the last three are the practice the engineer iterates indefinitely** (see §3 for the full presentation and §4 for why the practice is structurally a cycle).

- **P1 — Substrate.** Internalise the LLM as a software component: the structural framing (`tools → system → messages`) under current provider prompt-caching schemes, the four practical constraints (cost, hallucination, context, time), and the cache discipline that follows from them.
- **P2 — Building blocks.** Vocabulary fluency: function calling, MCP versus CLI versus the liteshell pattern, the agent loop, skills, characters, scaffolding.

- **P3 — Prototype with a general-purpose agent.** Use Claude Code, OpenCode, or Cursor as pair-programmer and reconnaissance tool to build the first working prototype against the real platform.
- **P4 — Harvest, fold, ship as CLI.** Collect tools, security scaffolding, skills, characters, and instruction sets discovered in P3; fold them into a small agent loop; ship the result as a command-line interface — *the Turtle pattern*.
- **P5 — Agent-tests-agent.** Use a general-purpose CLI agent to drive the custom agent through scenarios; treat behavioural evaluation as a complement to (not a replacement for) classical software-engineering testing.

The title — *Agents All the Way Down* — names the recursive structure of P3–P4–P5: the agent the engineer builds is itself a tool that another agent can drive. The general-purpose agent that drove the prototype (P3) is the same agent that tests the shipped artifact (P5). The orchestrator that composes specialists (§5) is itself an agent. The image is borrowed from the cosmological-regress anecdote popularised in modern English by Hawking [Hawking 1988], who attributes it to a lecture by Bertrand Russell. The methodology takes the image seriously and structural-recursively, not paradoxically.

### 1.4 Contributions

This paper makes four contributions:

1. **A named end-to-end methodology in five phases**: substrate, building blocks, prototype, ship-as-CLI, agent-tests-agent. To our knowledge no current publication chains all five into a single practice (see §2). Each phase has explicit acceptance criteria and an anti-mandate list.

2. **A novel quality-assurance discipline (P5): agent-tests-agent**. Classical software-engineering testing provides typing, unit tests, integration tests, and end-to-end tests; all four address deterministic behaviour and are necessary but not sufficient for agents, whose surface includes stochastic, scenario-shaped behaviour. We frame agent-tests-agent as a complementary behavioural-evaluation layer: a general-purpose agent drives the custom agent through scenarios, observes outputs and tool traces, and returns a structured evaluation. Classical testing and agent-tests-agent address different failure modes and are best deployed together.

3. **A multi-agent orchestration model derivable from the methodology itself**: once the agent is shipped as a CLI (P4), inter-agent composition is just CLI composition. We name this *the Turtle corollary* and observe that established orchestration patterns — supervisor/worker, parallel fan-out, conditional handoff, Ralph loops [Huntley 2026] — emerge naturally without framework primitives.

4. **An argument — not a demonstrated result — for dependency-aware agent design as a security posture**. The supply-chain incident record of 2024–2026 (CVE-2024-3094 in xz, the 2026 PyPI LiteLLM/Telnyx incident, sustained PyPI malware volume) makes dependency risk a routine engineering concern, and a custom agent built by this methodology has, by construction, a small dependency surface — all else equal, structurally less exposed than a framework-stack equivalent. We present this as a *posture*, deliberately of lesser evidentiary weight than contributions (1)–(3): its empirical validation — measured dependency count and attack-surface comparison across matched stacks — is left to future work (§7.2).

### 1.5 Scope and audience

This is a methodology paper: its subject is the practice of building custom agents, end to end. The frameworks surveyed in §2 are there to locate that contribution, not to rank them, and the AAC and LAMB appear as a worked example whose lessons hold outside the EdTech domain they came from (§7.5).

It assumes a competent practitioner who has used at least one LLM provider API, function calling, and one general-purpose code agent such as Claude Code. You need no framework experience and no Python — the examples are in Python only because it is convenient.

### 1.6 Paper map

§2 surveys the published agent-construction literature and locates this paper's contribution. §3 presents the five phases in sequence, with the preconditions / practice asymmetry made explicit at the opener; the framework-free design choice is also explained there. §4 promotes the P3 → P4 → P5 cycle from addendum to centre — after preconditions are in hand, the methodology *is* the cycle. §5 develops the Turtle corollary — orchestration as CLI composition — and illustrates the multi-agent pattern. §6 grounds the methodology in the AAC build as internal deployment evidence, with the AAC in production at two universities since April 2026. §7 discusses when frameworks remain the right choice, the supply-chain security argument for dependency-aware design, the build / deploy memory asymmetry that separates Turtles from Splinters, and open empirical questions. §8 concludes. **Appendix A** lists acronyms and methodology-specific vocabulary; the reader unfamiliar with any of the agent-engineering or software-engineering terminology may keep it open alongside the body.

---

## 2. Background: AI agent construction, 2022–2026

The published history of LLM-driven agents is short. We sketch it in four beats to locate this paper's contribution.

### 2.1 The chatbot era (2022 – mid-2023)

The release of ChatGPT in November 2022 [OpenAI 2022] helped make the conversational LLM interface a mass-market product surface. The early "agents" of this period were prompt-chaining demos — multi-step pipelines orchestrated by hand-written Python, with the LLM as a text-in / text-out component. The ReAct paper [Yao et al. 2022] articulated reasoning-and-acting as a pattern but appeared before mainstream structured tool-calling APIs were widely available, so it framed the loop conceptually before later API standardisation made such loops substantially easier to implement in production settings.

### 2.2 The function-calling era (mid-2023 – 2024)

OpenAI's introduction of function calling in June 2023 [OpenAI 2023] turned tool invocation into a first-class API interaction pattern. The LLM could now request a tool by emitting a structured object; the host program executed; the result returned as a tool message. Anthropic, Google, and the open-weight providers followed within months. Function calling became the fundamental agent cycle: prompt → tool call → execution → result → next prompt.

LangChain [LangChain 2026] was one of the most visible early orchestration libraries during the first wave of tool-calling applications. Its strength was breadth — every LLM provider, every vector database, every external tool wrapped behind a unified API. Its retrospective weakness, as reported by practitioners across blogs and conference talks during 2024–2025, was the cost of those abstractions: failure modes were difficult to diagnose through the framework's layers.

### 2.3 Framework proliferation (2024 – 2025)

The 2024–2025 period saw the rise of multiple orchestration styles, including graph-based (LangGraph [LangChain 2026b]), crew-based (CrewAI [CrewAI Inc. 2026]), multi-agent conversation (AutoGen [Wu et al. 2023]), declarative compilation (DSPy [Khattab et al. 2023]), and others. Each encoded a distinct opinion about what multi-agent orchestration should look like.

Concurrently, the academic community produced methodological precursors that anticipate individual pieces of the practice this paper names. Reflexion [Shinn et al. 2023] articulates agent self-reflection — the closest published prior art to P5 (agent-tests-agent). Self-Refine [Madaan et al. 2023] explores iterative-refinement patterns. Toolformer [Schick et al. 2023] grounds tool use. Voyager [Wang et al. 2023] grounds skill accumulation. The 2023–2025 survey literature [Wang et al. 2023b; Xi et al. 2023] documents the explosion of agent systems but does not propose a unified construction methodology. The same span also saw a popular **autonomous-agent wave** — AutoGPT and BabyAGI [Nakajima 2023] — which, though not peer-reviewed, was widely influential in popularising the idea of an LLM decomposing and pursuing its own task loop; it belongs in an honest chronology even though neither offered a reusable construction practice. What this paper contributes is a single sequential framing that links such ingredients into a five-phase engineering workflow.

### 2.4 The code-agent era (2024 – present)

The release of Claude Code [Anthropic 2026a], followed by OpenCode [Anomaly 2026], the maturation of Cursor [Cursor 2026] and Aider [Gauthier 2026], and the spread of repository-aware coding agents through GitHub Copilot and similar products, established a new tier of artifact: the **general-purpose code agent**. These products increasingly expose repository-aware editing, terminal access, and iterative tool use in production settings.

The code-agent tier did three things to the agent-construction landscape:

- **It demonstrated that custom agent loops were tractable.** Anthropic's "Building Effective AI Agents" (December 2024) [Anthropic 2024] documents a small-loop pattern in which most additional complexity lives in dispatch, persistence, and security scaffolding rather than in framework abstractions; in the same period HuggingFace's SmolAgents [HuggingFace 2024] pushed the complementary *tools-as-code* (code-as-action) approach, collapsing the tool surface to a single execution path.
- **It gave practitioners a tool for prototyping their own agents.** A code agent driving the engineer through a custom agent build is the dynamic this paper's P3 phase names.
- **It exposed the methodological gap.** Practitioners with access to Claude Code or OpenCode could *build* custom agents productively, but the *practice* of building them remained scattered across vendor docs, blog posts, podcast appearances, and leaked system prompts.

### 2.5 The methodological gap

Across the four eras, the agent-construction literature has converged on a recognisable vocabulary (function calling, MCP, the agent loop, skills, evaluation) and a set of orthogonal contributions: ReAct for the reasoning loop, Reflexion for self-evaluation, Toolformer for tool use, Voyager for skill accumulation, AutoGen for multi-agent conversation, LangGraph for graph orchestration. What is missing is a **named, sequential practice** that chains these contributions into a single transferable methodology — across one phase taxonomy, with explicit acceptance criteria, framework-independent. The closest existing documents — Anthropic's engineering blog, the MCP specification, the various framework READMEs — cover one or two phases each; none cover all five.

The gap is not that the pieces are missing — it is that no published work chains them. The contributions each cover a fragment of the path and stop: Anthropic's engineering writing gives substrate intuitions and an effective-agents framing [Anthropic 2024]; function-calling and MCP documentation give the tool-use mechanics [Anthropic 2024b; MCP 2025]; the research line on self-evaluation, tool use, and skill accumulation each formalises a single mechanism (Reflexion [Shinn et al. 2023], Toolformer [Schick et al. 2023], Voyager [Wang et al. 2023]); the orchestration frameworks (LangGraph, AutoGen, CrewAI) give multi-agent coordination; and the surveys catalogue the field without prescribing a build [Wang et al. 2023b; Sapkota et al. 2025], the latter distinguishing "AI Agents" from "Agentic AI" as concepts without proposing a construction methodology. Each covers one or two phases; **none presents substrate → building blocks → prototype → ship → test as a single named, framework-independent practice with explicit acceptance criteria.** That an emerging *agentic software engineering* discipline is now framing modeling, development, testing, deployment, and orchestration as one combined area [IEEE Software 2026] signals the same gap from the other side. The contribution of this paper is to fill it, framework-free by construction.

---

## 3. The methodology — five phases

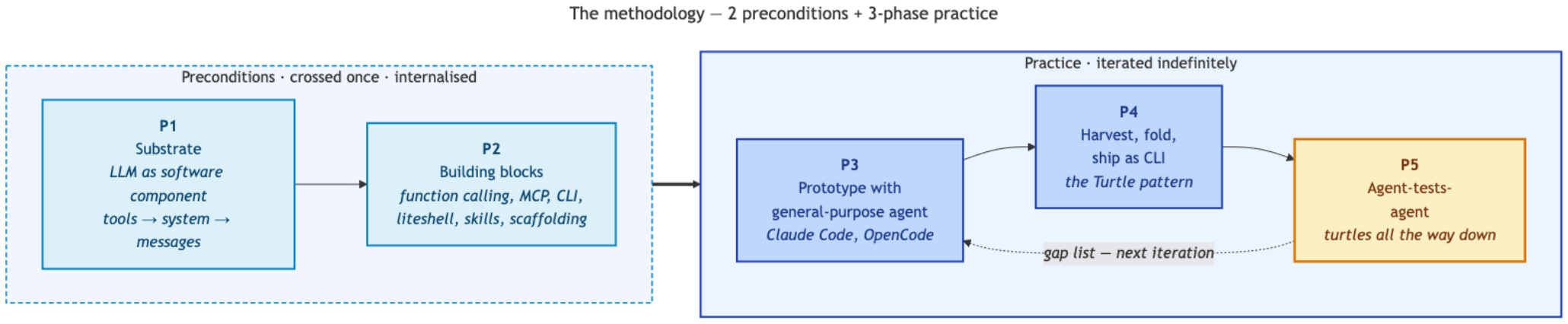

Figure 1: Figure 1. *The methodology in one diagram* — five phases, two-plus-three: P1 (substrate) and P2 (building blocks) are preconditions the engineer crosses once and keeps; P3 (prototype), P4 (ship as CLI), and P5 (agent-tests-agent) are the practice the engineer iterates indefinitely. The dashed arrow from P5 to P3 is the iteration spine (§4) — the practice loop the engineer lives in after preconditions are in hand.

**Preconditions and practice — the structural asymmetry of the methodology.** The five phases are presented in strict sequence, but they are not five iterations of the same kind of work. **P1 (substrate) and P2 (building blocks) are *preconditions*** — what the engineer must internalise before the methodology can be applied. Their acceptance criteria are *knowledge*: the engineer can read a usage record and diagnose its cost shape (P1); the engineer can name each building block and pick the right one (P2). P1 and P2 are crossed once and kept; they are not iterated. **P3 (prototype), P4 (ship as CLI), and P5 (agent-tests-agent) are the *practice* itself.** Their acceptance criteria are *artifacts*: a working prototype, a CLI invocation, a scenario suite. The engineer iterates P3 → P4 → P5 → P3 indefinitely once preconditions are in hand. **The methodology's recursion — *agents all the way down* — lives in the P3-P5 loop.** P1 and P2 are the floor on which the loop stands. The presentation order below preserves the strict sequence, but the reader should hold the asymmetry: P1 and P2 are read once; P3–P5 are read with the recognition that this is where the engineer's working life happens.

**On frameworks.** The methodology is framework-free by construction. §7.1 sets out the engineering reasons — churn-prone abstractions, a larger dependency surface (§7.2) — and the cases where a framework is still the right call.

**Phase structure.** Each phase below has (i) a substantive claim, (ii) explicit acceptance criteria, and (iii) anti-mandates — things the practitioner should *not* do. The sequence matters because each phase's correctness depends on choices made under the previous phase's framing.

### 3.1 P1 — Substrate

**Claim.** The engineer building a custom agent must internalise the LLM as a software component. This is more than *use the LLM API correctly*. It means keeping five things in view at once:

- how next-token prediction actually works;
- how the tokeniser shapes both cost and behaviour;
- the fixed-size context window as a resource every part of the prompt competes for;
- the alignment stack's role in setting the model's defaults;
- the four practical constraints — cost, hallucination, context, time.

Without this internalisation, every later decision is taken in the dark.

**A structural framing.** A substantive addition to the standard LLM-as-component framing: for Anthropic-style prompt caching [Anthropic 2026b], the cacheable prefix is ordered as `tools`, then `system`, then `messages`:

```
[ tools ]    ← declared at agent init; most stable; cached longest
[ system ]   ← set at session start; stable per session; cached
[ messages ] ← grow per turn; only the latest are fresh tokens
```

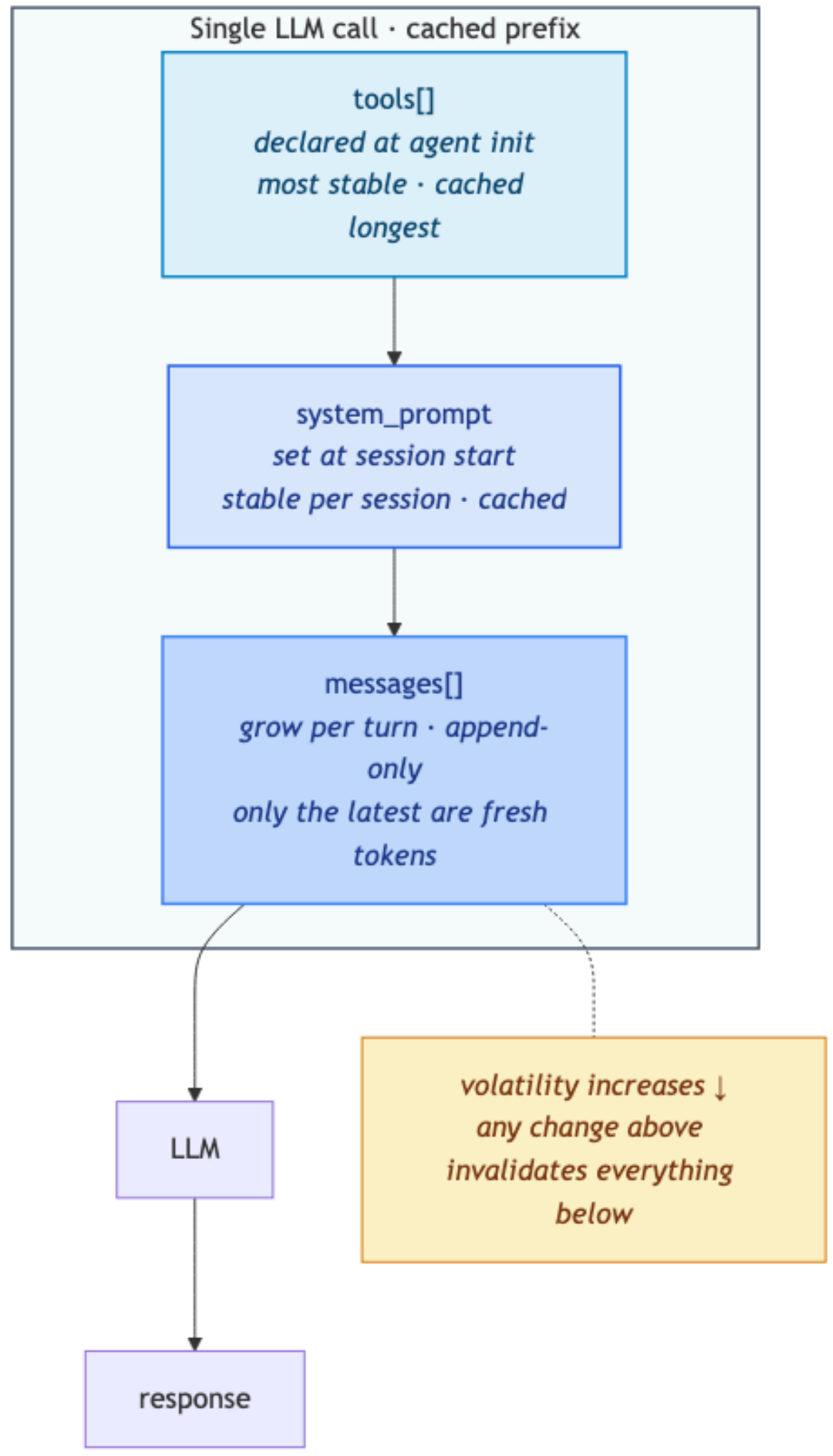

Figure 2: Figure 2. *The cached prefix* — `tools → system → messages` as a single layered structure handed to the LLM on every call. Volatility increases top-to-bottom; any change above invalidates the cache for everything below.

Stable content placed earlier in that prefix is cheaper to reuse across turns. The discipline is to segregate volatility so the stable parts remain cached; volatility is not the same as importance (messages are the most volatile and also where the work happens; the lesson is about structure, not priority). Any change to `tools` or `system` invalidates the cache for everything downstream — reordering tools alphabetically as a "cleanup" can erase a session's accumulated cache.

**Cache pricing in practice.** Current official pricing pages show that cache-read or explicit context-caching discounts are often close to one tenth of base input cost on Anthropic [Anthropic 2026b], on OpenAI flagship text models [OpenAI 2026], and on Google Gemini entries where explicit caching is available [Google 2026a; Google 2026b]. DeepSeek's published cache-hit pricing [DeepSeek 2026a; DeepSeek 2026b] is substantially lower — model-specific cache-hit multipliers approach 0.02× and below at current rates. Google's explicit caching additionally bills storage. **Cache discounts are provider- and model-specific, often substantial, and should be cited directly from current provider documentation when reported.** The engineering implication is invariant across providers: engineering for cache hits makes the agent cheap and fast by design, not by luck.

**Key-Value (KV) cache discipline.** Tools are declared once at init and immutable during the session. System is set once at session start and holds five categories of content: one-shot tool-use examples, worldview / domain knowledge, persona, directives, memories. Messages grow strictly append-only — never rewritten in place, never spliced, never popped. Per-turn dynamic context (timestamps, the user's current question, tool results) belongs in messages, not in system: placing stable instructions in the cached prefix and keeping volatile per-turn state in message history rather than the system block is the recommended posture [Anthropic 2026b].

**Acceptance.** The engineer can read an LLM call's usage record and diagnose its cost shape on first read. Given a working agent, can identify where its cache prefix begins and ends and why each section is positioned where it is.

**Anti-mandate.** Do not put dynamic context (current timestamps, runtime state, the user's current message) into the system prompt for cache hygiene. Do not reorder `tools` or `system` between sessions if no functional change is needed.

### 3.2 P2 — Building blocks

P2 is the **vocabulary** the methodology operates in. Each block is a tool in the toolkit, not a topic in its own right. The goal is to make each block legible enough that P3 and P4 can pick the right one off the shelf deliberately.

**Function calling.** The fundamental agent cycle: pass the model a list of tool schemas (`{name, description, parameters}`); when the model decides a tool is needed, it returns a structured tool-call object; the host code intercepts, executes, and returns the result as a tool-result message. Three operational concerns. First, the *tool-count problem*: large tool catalogs can degrade focus and increase prompt overhead [Anthropic 2024]; numeric thresholds for "too many" should be quantified empirically per use case rather than relied on as folklore. Second, *schema design*: atomic operations, documentation-quality descriptions, JSON-Schema-clean parameters, graceful failure modes. Third, *where security lives*: never in the schema, never in the description, never in the system prompt, but in the scaffolding's tool dispatcher — prompt instructions help steer behaviour, but authorisation and execution policy belong in deterministic host-side controls.

**MCP, CLIs, and the *liteshell* pattern.** The Model Context Protocol [Anthropic 2024b; MCP 2025] is a useful open standard for connecting LLM applications to tools and external context. In local engineering workflows, however, CLI surfaces often remain attractive because models are heavily exposed to shell documentation and usage patterns during pretraining, and because credential-handling differs by transport. The current MCP authorisation specification [MCP 2026] covers HTTP-based transports; STDIO transports are expected to read credentials from the environment. That is a narrow, transport-specific fact about local credential handling. We do not read it as evidence that OAuth-over-MCP is immature across the board.

**A cost model, not a single ratio.** The token cost of either surface decomposes cleanly, and the comparison is better stated as a calculation than as a headline number. An MCP call pays, *per session*, for the **tool registry** — every exposed tool's name, description, and JSON schema, fixed at the head of the conversation — and then, *per call*, for structured arguments and a structured result the model must take into context whole. A CLI call pays for the command text and the command's stdout / stderr, with no per-session registry:

```
MCP cost ≈ tool registry + schema + arguments + structured result + reasoning per
call
CLI cost ≈ command + terminal output + reasoning per command
```

The governing quantity is therefore not "MCP versus CLI" in the abstract but overhead ≈ (number of exposed tools) × (size of each tool's description and schema) × (verbosity of returned data).

The MCP specification makes the registry explicit — tools are named operations with descriptions, typed inputs, and metadata exposed to the model [MCP 2025] — and Anthropic's own engineering guidance notes that connecting many MCP servers creates enough tool context to motivate code-execution patterns that reduce it [Anthropic 2026c]. The same variables bound the CLI side too: a CLI that dumps huge raw output (verbose logs, untrimmed JSON, stack traces) can cost *more* than a well-scoped MCP tool. The engineering discipline is to keep tool registries small and outputs trimmed, on whichever surface — but the CLI starts from a structurally lower baseline because it carries no registry.

Reading the published reports through that model rather than as one number yields the following shape:

| Workload | MCP token cost vs a lean CLI call | What drives it |
|---|---|---|
| Single well-scoped tool, atomic action | ≈ 1.2×–2× | one tool's schema + one structured result |
| Multi-step task, several tools loaded | ≈ 2×–6× | a registry of many tool descriptions + per-step JSON |
| Large / noisy tool registry | > 10× | registry tokens dominate the context window |

| Workload | MCP token cost vs a lean CLI call | What drives it |
|---|---|---|
| CLI with trimmed output | baseline | command text + concise stdout |
| CLI with huge raw output | can exceed MCP | undisciplined stdout (logs, dumps, stack traces) |

*Table 1: MCP-versus-CLI token overhead as a function of tool count, schema verbosity, and result size — not a fixed ratio. The bands are indicative (reasoned from the cost model above); the measured data points are the three benchmarks cited below.*

The headline practitioner benchmarks sit at the high end of this range, exactly as their conditions predict: a controlled matched-task benchmark reports ≈35× more tokens via MCP than via CLI (with task-completion reliability dropping from 100% to 72% on harder scenarios) [MindStudio 2026]; a single GitHub language-check costs ~1,365 tokens via `gh` against ~44,026 via the matching MCP server [Vensas 2026]; and a GitHub MCP server exposing 93 tools adds ~55,000 tokens of registry overhead at session start versus ~200 for the `gh` equivalent [Reinhard 2026]. These are large-catalogue, many-call measurements — we report them as the upper end of the model above, not as a universal ratio. The directional claim is invariant across the whole range: MCP carries fixed-and-variable token overhead that CLI does not, and that overhead **compounds across repeated agent calls**.

Token volume is not the only axis, and arguably not the deepest one. A second, qualitative difference is **composability**. An MCP tool returns a fixed result structure that the model must take into context *whole* and then filter by reasoning over it; a CLI returns a stream that the model can compose with the shell's own filters — piping through `jq`, `grep`, `head`, or a one-off script — so that the model narrows the result *before* the bytes ever enter the context window. The practitioner argument is that this makes context pollution the default failure mode of MCP and an avoidable one under CLI: *"as a model I always have to get the huge blob back… but if I would build the same as a CLI… it could just add a JQ command and filter itself… you have no context pollution"* [Steinberger 2026]. The same source notes the historical value MCP nonetheless delivered — it pushed providers to expose APIs that a CLI can now be written over — and concedes the case where MCP's stateful session model genuinely fits, the canonical example being browser automation [Steinberger 2026]. We treat that as the boundary condition, not the refutation: CLI is the default surface; MCP earns its place where a tool requires sustained stateful interaction the shell idiom does not model cleanly.

**The *liteshell* pattern — a name for a convergent practice.** When the custom agent runs inside a cloud application — FastAPI, Spring Boot, Django, Express — there is no host shell to invoke. Engineers building custom agents through 2024–2026 have been convergently rediscovering the same workaround, and it is an old one with a name: the **Facade** [Gamma et al. 1994] — a single, simplified interface placed in front of a complex subsystem. A liteshell *is* that Facade, with two specifics that make it worth its own name: the simplified interface it presents is **CLI-shaped**, and the consumer it presents it to is **the LLM**. Concretely: build a small in-process facade that *presents itself to the LLM as if it were a CLI*, then dispatch each "invocation" to a Python method or a loopback call against the application's own REST API. The model sees `call_cli tickets list --status=open`; the host parses the string, dispatches, returns the result as a tool message. The shape is one tool in `tools[]` (the dispatch tool), one auth setup at library init, and one short skill (markdown) with two one-shot examples — enough for the model to extrapolate the rest from its pretraining exposure to shell idioms.

**The contribution here is naming, not invention** — *we did not invent these; we named them*, in the lineage of [Gamma et al. 1994], where *Design Patterns* did not invent the patterns it described but gave them a vocabulary that thirty years of software engineering then borrowed. We give this pattern the name **liteshell** because, to our knowledge, no widely-adopted name exists for it, and its underlying components are decidedly not novel: the Facade pattern is thirty years old [Gamma et al. 1994]; code-execution tools that give the LLM a single dispatch surface — Anthropic's Computer Use with its bash and computer tools [Anthropic 2024c], HuggingFace's SmolAgents with the tools-as-code pattern [HuggingFace 2024] — predate this paper and embody the same structural move (collapse the tool surface to one execution tool; let the model author what to invoke); the Anthropic Skills format [Anthropic 2025] provides the markdown-skill convention we use to teach the model the invocation grammar; and GitHub's `gh` and Google's `gws` are *external* liteshells the engineering community has already standardised on, distributed as packaged binaries rather than as a class inside the engineer's own application. **What this paper adds is the name, the locating, and the worked example.**

**Why the name earns its place.** The liteshell carries the CLI-over-MCP token argument into the cloud-runtime case. The benchmarks above measured *external* CLIs (`gh`, `gws`); the liteshell extends the same advantage to where no host shell exists, because the model still treats the in-process facade as a CLI and applies the same pre-trained shell knowledge. Naming the pattern makes it a deliberate engineering choice rather than an accident of implementation.

**The agent loop.** A minimal tool-calling loop can be implemented directly in a small amount of host-language code, with most additional complexity living in dispatch, persistence, and security scaffolding rather than the loop body itself [Anthropic 2024]. The structure:

```
messages = []
while True:
    response = llm.chat(
        tools=tools,            # static, declared at init
        system=system_prompt,   # static, set at session start
        messages=messages,      # grows append-only
```

```
    )
    if not response.tool_calls:
        return response.content
    for tc in response.tool_calls:
        result = dispatch(tc)            # scaffolding's responsibility
        messages.append(assistant_turn(tc))
        messages.append(tool_result(result))
```

*Listing 1: A minimal agent loop. The boundary does real work — the LLM thinks and chooses; the scaffolding does everything else.*

**Skills, characters, instruction sets.** The content layer of P2, living in the system prompt. *Skills* are markdown files describing how to do a specific thing — loaded into the system prompt at session start or on demand, editable without changing code. *Characters* are voice / stance / register — they constrain which outputs the model considers plausible. *Instruction sets* are rules — the must-dos and must-not-dos in the system prompt's directives section. Discipline: do not put security in the instruction set. Instructions shape *behaviour*; the scaffolding's allow-list shapes *outcomes*.

**Hooks.** The deterministic counterpart to the content layer. A *hook* is a host-side function that fires on a fixed point in the agent's lifecycle — most importantly **before a tool call executes** (`pre-tool-use`), but also on session start, on tool result, or before output is returned. Where instructions and skills *steer* the model probabilistically, a hook *enforces* outcomes in ordinary code: it is where the allow-list is actually checked, where a requested `call_cli` is admitted or denied, where a skill or a tool argument can be scanned before it is trusted (a deployed instance of this pattern scans every contributed skill through a malware-detection service before the agent may load it [Steinberger 2026]). Hooks are the mechanism that makes the §3.2 discipline — *security lives in the scaffolding, not the prompt* — concrete: the instruction set expresses intent, the hook is the gate. P4 (§3.4) shows the `pre-tool-use` hook folded into the loop.

**Acceptance.** The engineer can name each of these blocks and pick the right one for a given problem. Has implemented at least one custom tool with proper schema; has driven the agent loop manually (without a framework) for at least one toy problem.

**Anti-mandate.** Adopting a framework before this vocabulary is internalised is the most common path to opaque failure modes later. Frameworks remain a defensible engineering choice once the building blocks are in hand (see §7); before that, framework adoption typically substitutes API fluency for principle fluency.

### 3.3 P3 — Prototype with a general-purpose agent

P3 is where the methodology becomes a practice rather than a body of knowledge. The move: **use a general-purpose agent — Claude Code, OpenCode, Cursor — as a pair-programming partner and reconnaissance tool**, with the explicit goal of building a custom agent for a specific application.

**The reconnaissance step.** Before any custom agent code is written, the general-purpose agent surveys the platform the engineer is building inside. For the AAC build (see §6), this was the first session: Claude Code walked LAMB through its existing `lamb-cli` tool — `lamb assistant list`, `lamb model list`, `lamb kb list`. The first gap discovered was that no `lamb rubric` command existed, even though assistants referenced rubrics. **The reconnaissance step exposes what is missing before the custom-agent design begins.**

**Architecture-discovered-by-building.** The classical software-engineering move is to design the architecture first, then build to specification. The methodology's move is to **build first** and let the architecture emerge from what works and what does not. Three things engineers commonly discover this way: where their data actually lives (often not where the design diagram says); what tool surface the LLM can drive cleanly (often coarser than the platform's API surface); and where the security boundary belongs (almost always at the scaffolding's tool dispatch, not in the prompt). The reconnaissance pass produces a working prototype, a list of needed tools, a first draft of the system prompt, and a list of gaps.

**Tools and security in P3.** Tool dispatch is in host code, not in the model's text. Allow-list everything — the model can *request* `call_cli ls /tmp`; the host code decides whether `ls /tmp` is allowed.

A note on secrets: **never expose a secret to the model, never persist one in a prompt, never log one in plaintext. When the agent needs a credential, fetch it from an approved secret store at runtime and pass it only over an authenticated, secure channel.** A common manifestation is operating-system keychain integration: the credential lives in the keychain, the runtime fetches it on demand, and it never appears as a string in tool arguments, the system prompt, or model output. These are not implementation details; they are the methodology's security contract, established in P3 and inherited by P4.

**Acceptance.** A working prototype demonstrates the agent's core loop against the real platform. A list of tools — pre-existing CLIs and new liteshell entries — is in hand. A first draft of the system prompt with one or two skills loaded exists. A documented list of gaps (missing APIs, security boundary questions, ambiguous authorisation cases) is recorded.

**Anti-mandate.** Do not skip P3 because the architecture is *already known*. It is not. The architecture emerges from the build; the few hours of reconnaissance save days of P4 rework.

### 3.4 P4 — Harvest, fold, ship as CLI (the Turtle pattern)

P4 is the engineering phase: take what was discovered in P3 and produce the production agent. The phase has three movements: *harvest*, *fold*, and *ship as CLI*.

**Harvest.** From P3's working prototype: tools (schemas + implementations, pulled into a clean `tools/` directory); security scaffolding (the allow-list, secret-retrieval setup, authorisation policy); skills (promoted from loose snippets to first-class version-controlled markdown files); characters (promoted from experimental phrasing to explicit persona content); instruction sets (rules, the directives section). The prototype's system prompt is usually a mess of inline instructions; harvest is a deliberate collection step.

**Fold into a small loop.** The reference shape generalises Listing 1 with two additions, both promoted from prototype to first-class: a **session store** that persists the growing `messages[]` list between turns under a session id, and a **pre-tool-use hook** — the explicit security check that runs in host code on every tool call the model emits, *before* the tool actually executes (the *Hooks* block of §3.2, now doing its real work). Listing 2 shows the hook folded into the loop: the only change to the body is that `dispatch` is now gated by the hook, which admits or denies the call deterministically before any side effect occurs. Everything else is the surrounding state-and-policy machinery that graduates from *"the prototype's shortcut"* to *"the production agent's discipline"*. The cache discipline from P1 is preserved by construction: tools immutable at instantiation, system immutable per session, messages append-only.

```
for tc in response.tool_calls:
    if pre_tool_use_hook(tc):           # deterministic security gate
        result = dispatch(tc)           # allowed → execute
    else:
        result = denied(tc)             # blocked before any side effect
    messages.append(assistant_turn(tc))
    messages.append(tool_result(result))
```

*Listing 2: the `pre-tool-use` hook folded into the loop body of Listing 1. The agent still proposes the call; the hook — ordinary host code, not a model decision — decides whether it runs. This is where the allow-list lives, and why authorisation never belongs in the prompt.*

**Ship as a CLI tool — the Turtle pattern.**

The defining move of P4: **the agent is itself a CLI tool**. Not a chat window in a browser, not a hidden endpoint on a backend, not an SDK to be imported elsewhere. A command-line program with a stable invocation shape:

```
turtle "do thing"
  → response + new-session-id
turtle <session-id> "do next thing"
  → continues the session
turtle --list-sessions
turtle <session-id> --replay
  → re-runs the session (traceable; reproducibility caveat below)
```

*Listing 3: the Turtle-pattern CLI invocation shape — stateless calls, session-keyed continuation, replay, and `--help` for the next agent.*

We name this *the Turtle pattern*. The name points, deliberately, to three things at once. The proximate reference is the *turtles all the way down* image that gives the methodology its name and that anchors P5 (§3.5). The deeper reference is Seymour Papert's Logo turtle [Papert 1980], the canonical educational-programming primitive — children's first encounter with autonomous symbolic action, and the historical antecedent of the agent-as-instructable-actor framing. The third is more practical: a CLI is, by construction, a tool that another agent can drive.

The agent has:

- **Stateless invocation** per call (the session store is the state-holding surface).
- **Session-id-keyed continuation** for multi-turn use.
- **Structured outputs** — the response itself + the session id + optionally a JSON sidecar with tool traces.
- **`--help` that works** — both for human readers and for the agent driver in P5.

A reproducibility caveat: the `--replay` flag should be framed as **traceable or approximately reproducible** rather than as a deterministic re-run guarantee. Bit-exact determinism across LLM invocations requires provider-specific guarantees, fixed seeds where supported, and matching runtime versions; in practice replay is most useful for inspection, diff-against-rerun, and behavioural debugging rather than as a regression-test substitute.

The design constraint: *the agent must be built so that it is friendly for the **other agent** to drive*. A chat-window-in-a-corner deployment is not testable by another agent. A CLI is. This single discipline — *built for the next agent to drive* — shapes the I/O surface choices that make P5 tractable.

**Acceptance.** The agent has a CLI invocation that returns reliably. Sessions persist, can be resumed, can be replayed (as a trace, not as a bit-exact rerun). Tools, system prompt, skills, characters live in version-controlled files (markdown for skills/characters; host-language code for tools; structured config for security). Security is enforced in the scaffolding's tool dispatcher, not in the prompt. The CLI's output is parseable enough that a general-purpose agent can drive it — see §3.5.

**Anti-mandate.** Do not put authorisation logic in the prompt. (This is the AAC build's most-repeated lesson — see §6.) Do not ship without `--help` working; that removes the next agent's onboarding surface.

### 3.5 P5 — Agent-tests-agent

P5 is the methodology's novel piece. Classical software-engineering testing — typing, unit tests, integration tests, end-to-end tests — was developed for deterministic software. Agents are stochastic; they have a behavioural surface that classical testing cannot exercise cleanly. Deterministic tests remain necessary for invariants, interfaces, and failure handling, but agentic behaviour also benefits from scenario-based behavioural evaluation. The closing image, again: *agents all the way down*. The custom agent is tested by another agent. That testing agent is itself driven by an agent. **The principle is recursion, not paradox.**

**The pattern.** A general-purpose CLI agent (Claude Code, OpenCode) is given three inputs: (i) a scenario — natural-language description of a task the custom agent should be able to handle; (ii) the custom agent's CLI invocation (the Turtle interface from §3.4); and (iii) an evaluation rubric describing what counts as a pass, what counts as a fail, what edge cases to probe.

The general-purpose agent then drives the custom agent through the scenario, observes outputs and tool traces, probes edge cases (adversarial inputs, ambiguous requests, multi-turn drift), and returns a structured evaluation: pass/fail per criterion, with evidence quotes from the transcript. The published Reflexion line [Shinn et al. 2023] anticipates the verbal-feedback aspect of this pattern; what is added here is naming the discipline, locating it in the methodology's sequence, and pairing it with a CLI artifact designed to be driven.

**What this gives.** Four properties classical testing cannot:

1. **Behavioural regression testing.** *"Does the agent still refuse to leak credentials when asked cleverly?"* is not a unit test; it is a P5 test.
2. **LLM-comparison testing.** Same scenario, different backbone — different providers, different model sizes — produces apples-to-apples behavioural comparison under a controlled scenario suite.
3. **Documentation generation as a side effect.** The evaluation agent's transcripts are example interactions; the good ones flow into the documentation.
4. **The improvement loop.** Failures reveal gaps; gaps go back into P3 as next-iteration prototypes; the methodology cycles (see §4).

§6 instantiates this discipline against the AAC build: driving the custom agent with a general-purpose agent surfaced a catalogue of behavioural defects that narrow developer tests had missed (Tables 3 and 4), each generalising to an anti-mandate — concrete evidence of what P5 catches that classical testing does not.

**The post-build refactor elicitation.** A lightweight, high-yield variant of the improvement loop runs at the *end* of every build or change, not only when a test fails. The engineer asks the general-purpose agent that just produced the work a deliberately retrospective question — *what would you refactor now that this exists?* — and acts on the friction it names. The premise is that some design defects are visible only once the artifact exists and has been exercised; eliciting them immediately, while the full working context is still loaded, is cheaper than rediscovering them on the next iteration. The same logic governs the engineer's own experience: a recurring point of friction in operating the agent — or in operating the surrounding toolchain — is itself a P5 signal, and the disciplined response is to spend the next iteration removing the friction at its source rather than working around it. This is established practitioner discipline: *"almost every time I merge a PR, build a feature, afterwards I ask, 'what can we refactor?' … because then you build it and you feel the pain points"* [Steinberger 2026].

**Classical testing sits alongside P5.** P5 does not replace classical testing. The two address different failure modes and are best treated as complementary. The full stack:

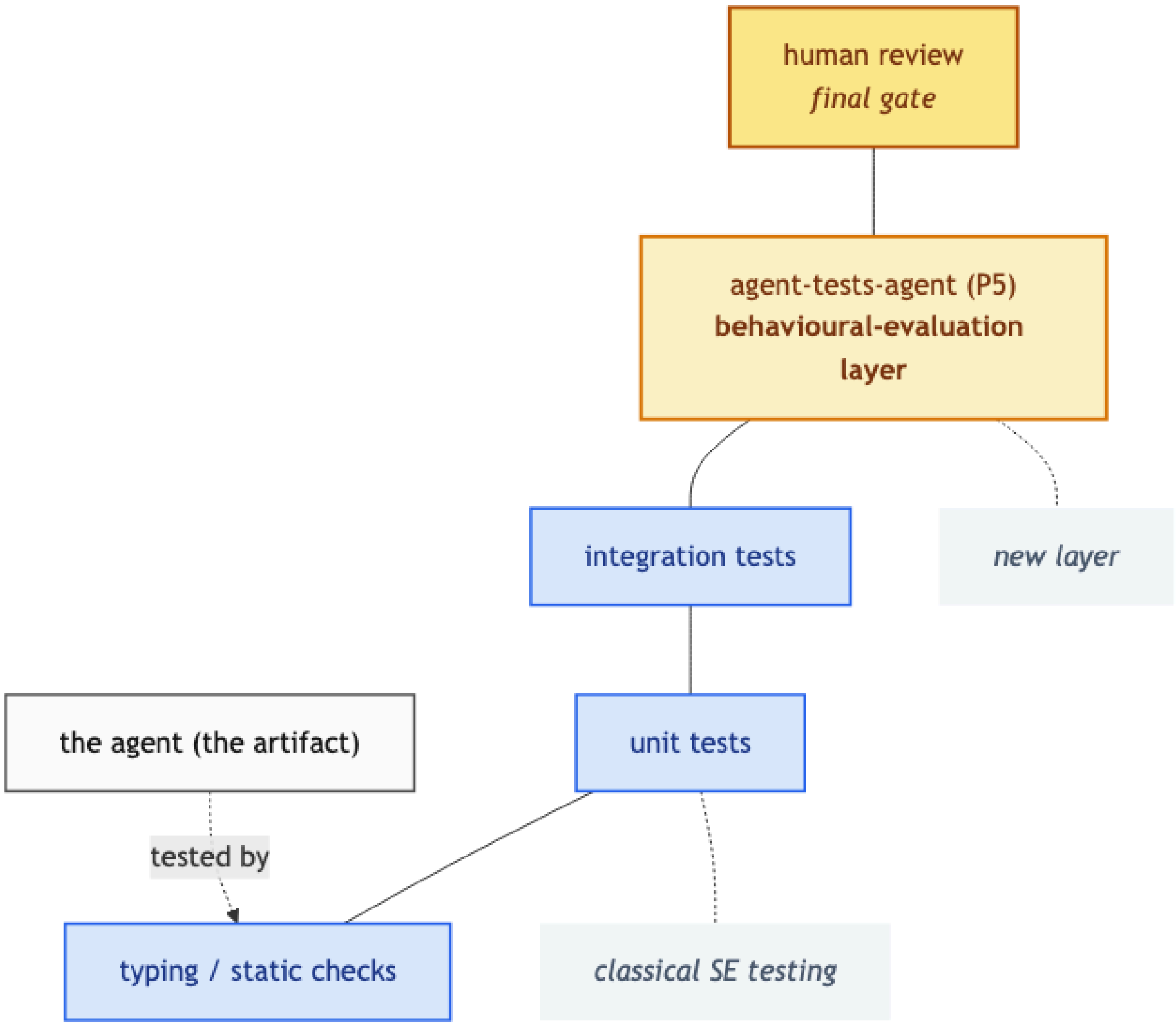

Figure 3: Figure 3. *The testing stack with P5 inserted between classical software-engineering testing and human review.*

The classical layers test that the plumbing is correct (deterministic behaviour, no exceptions, types match). The P5 layer tests that the *behaviour* is correct (the agent does what a user needs it to do, refuses what it should refuse, recovers from what it should recover from). For production deployments, both classes of test address different failure modes and are best deployed together.

**Acceptance.** A scenario suite covering the agent's primary responsibilities exists. A general-purpose CLI agent can drive the suite end-to-end without manual intervention. The evaluation output is structured (pass/fail per criterion, with evidence). At least two backbones have been tested against the same scenarios. The gap list from the most recent P5 run is the input to the next P3 iteration.

**Anti-mandate.** P5 is continuous: every change to tools, system prompt, skills, or character should re-run the scenario suite. Testing the tools individually is unit testing; agent-tests-agent is behavioural testing; they exercise different failure modes and are not substitutes.

---

## 4. The methodology is the cycle

Once preconditions P1 and P2 are in the engineer's hands, **the methodology is the cycle** P3 ⟶ P4 ⟶ P5 ⟶ P3 — and the engineer spends the rest of the agent's life inside it. §3 presented the phases sequentially because they have to be introduced sequentially; this section names what the phases become operationally, which is **a single cyclic practice with three named stations**.

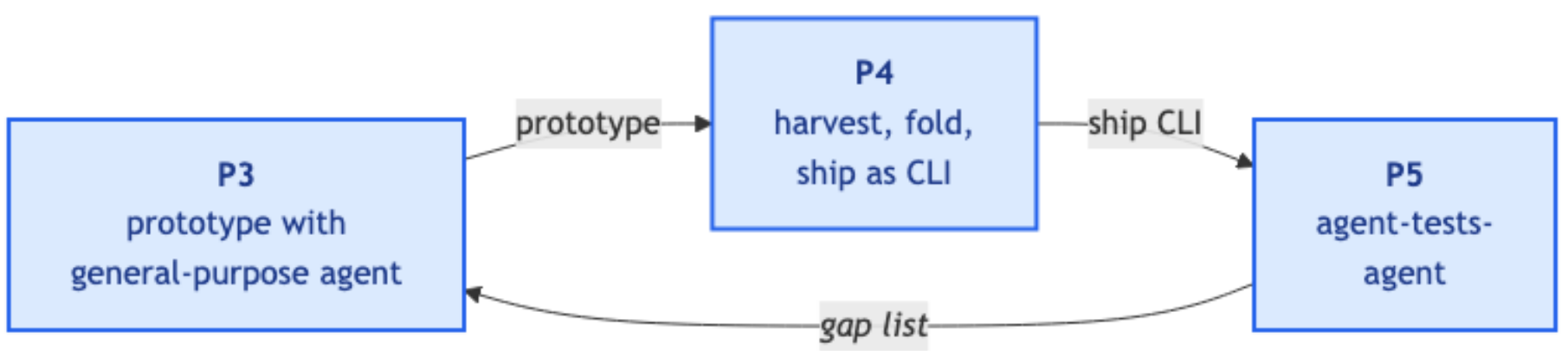

Figure 4: Figure 4. *The methodology after preconditions — the practice is the cycle P3 ⟶ P4 ⟶ P5 ⟶ P3. The engineer iterates indefinitely; P1 and P2 do not re-enter the loop because they were internalised once.*

**The structural property the cycle inherits from the preconditions.** The cycle is *cheap* because of P1. Most of the agent's prefix (tools + system) does not change between iterations, so cache hits dominate. Under current provider prompt-caching schemes (§3.1), each pass through the cycle is a small addition on top of the previous pass, not a rebuild from scratch. An engineer who iterates ten times in a quarter pays about the cost of one disciplined initial build plus ten thin refresh layers — not ten initial builds.

A back-of-envelope example makes the shape concrete. Take an agent with a stable prefix (tools + system) of ~20K tokens and ~2K tokens of fresh work per iteration. Without cache discipline, ten iterations re-pay the full ~22K each time: ~220K input tokens. With the prefix cached at the ≈0.1× cache-read rate typical of current providers (§3.1), the same ten iterations cost one full prefix (~20K), ten cached reads of it (10 × 20K × 0.1 ≈ 20K), and ten fresh deltas (10 × 2K ≈ 20K) — about **60K effective input tokens, under a third**. The numbers are illustrative; the direction is structural. The larger and more stable the prefix, the more the cycle rewards cache discipline.

**Why iteration matters substantively, not just economically.** In our experience building the AAC, each pass through the cycle made the agent measurably more capable: a single pass produced a less capable agent than three passes, and three passes a less capable agent than one re-tested every release. We report this as an observation from the construction period rather than a controlled result. The reason is that P5 (agent-tests-agent) is **the only place in the methodology that surfaces behavioural defects** — defects classical testing was not designed to detect, in a stochastic substrate that classical testing does not exercise. Each cycle is a behavioural-quality bump. **Skipping the cycle is choosing to ship the version of the agent that has been least exposed to other agents' attempts to break it.**

**Why this cycle remains tractable at production scale.** Cheap iteration on a transparent small loop survives the production calendar because the iteration cost does not include re-learning a framework's evolving abstractions. A framework refactor between minor versions (LangChain v0.x → v0.y, observed multiple times during 2024–2025) imposes an *iteration tax* on every team using the framework, independent of any change to the team's own agent. A framework-free cycle, built on stable primitives (function-calling APIs, MCP, CLI semantics), carries no such tax. The engineer's iteration cost is bounded by their own substrate, not by an upstream maintainer's release schedule.

**The over-orchestration trap.** There is a characteristic maturity arc in how engineers structure agentic work, and the methodology is calibrated to its endpoint. Newcomers begin with short, direct prompts; as they gain fluency they tend to over-build — elaborate multi-agent orchestration, long chains of bespoke sub-agents, large libraries of custom commands — on the assumption that sophistication scales with machinery; the most experienced practitioners then converge *back* on short prompts and a small, transparent loop, having found that the elaborate middle stage mostly adds failure surface. One practitioner names this middle stage the *agentic trap* [Steinberger 2026]. The P3 → P4 → P5 cycle deposits the engineer at the far end of that arc by construction: the deliverable of P4 is a single small loop, and the corollary of §5 derives multi-agent structure as plain CLI composition rather than as a dedicated orchestration layer. The trap is real and common; the methodology avoids it not by exhortation but by the shape of the practice.

**A note on stopping.** The methodology does not name an end-state. There is no "completed" agent under this practice; there are versions of the agent that have passed the most recent scenario suite under the current LLM backbone. *"Done"* is a release decision the engineer takes against acceptable behavioural-defect density on the latest P5 run, not a state the methodology promises. This is honest about the artifact's nature: a custom agent is a *living artifact*, like a service or a library that depends on an evolving runtime — its quality is maintained by ongoing iteration, not certified at a single point in time.

---

## 5. Orchestration as CLI composition — the Turtle corollary

The methodology produces, in P4, a custom agent shipped as a CLI. The agent has `call_cli` in its tool list, used in P3 and P4 to invoke the platform's existing CLIs (`gh`, `gws`, `lamb-cli`, the liteshell). It follows immediately that **another agent's CLI is just another tool**. There is no architectural distinction between:

- `call_cli gh issue list --label bug`
- `call_cli donatello "review this PR for security concerns"`
- `call_cli leonardo --strategy "decompose this multi-step task"`

All three are: parent emits a structured tool call → scaffolding dispatches → subprocess runs → output captured → result becomes a tool-result message in the parent's message stream. The parent does not know whether it called a Go binary, a bash script, or another LLM under the hood. **The CLI contract is the orchestration interface.**

Multi-agent orchestration in this methodology is therefore **not a phase** — it is a *corollary*. The agents built under P4 compose with one another because they compose the way every other CLI composes.

### 5.1 What the corollary gives

1. **Heterogeneous backbones.** A supervisor running on one provider's flagship can delegate to a research worker on a different provider, and a verification worker on a local model — the parent does not care; each child's CLI is the contract.
2. **Process isolation.** A child that crashes or hangs does not crash the parent — the parent sees a failed `call_cli` (non-zero exit or timeout) and recovers, and the child's memory is reclaimed by the operating system when it exits (see the shared-host caveat below).
3. **Independent deployability.** A new version of the writer agent replaces the old one without touching the orchestrator. The CLI surface is the API.
4. **Parallelism via the shell.** `turtle "X" & turtle "Y" & wait` — the primitive bash has provided for decades. The parent batches invocations and the scaffolding parallelises at dispatch.
5. **Composition with non-agent CLIs.** An orchestrator calls a Turtle calls `kubectl` calls another Turtle calls `gh`. Agents are not a special node type in the orchestration graph — they are CLIs with stochastic behaviour. Most frameworks silo agent-to-agent flow into one abstraction layer and tool-use into another; here they share one surface.
6. **Replayability and audit.** Each agent has a transcript. The parent's transcript references children by invocation. Grep, diff, replay — the trace surface is the same one used for single-agent debugging.

**A shared-host caveat — for (2) and (4).** Process isolation and shell parallelism are *fault*- and *composition*-level properties, not *resource*-level ones. When the orchestrator and its children share a host, memory, CPU, and file descriptors are common: a child that exhausts RAM can trigger the host out-of-memory killer. The kernel picks its victim by heuristics, and may kill the parent or an unrelated process rather than the offending child. Unbounded `&`-fan-out only multiplies the pressure. Resource isolation is therefore not free; the scaffolding must impose it, with per-child limits (`ulimit / setrlimit RLIMIT_AS`, a cgroup `memory.max`, a container, or a separate host) and a concurrency cap on parallel children. This is the same scaffolding discipline as the recursion-depth budget (§5.4): the substrate gives fault isolation; resource isolation is bought.

### 5.2 The Ninja Turtles — a multi-agent topology

A natural specialisation topology, illustrated using a four-character pattern that gave this paper's mascot its name:

| Agent | Role | What it does |
|---|---|---|
| **Splinter** | Orchestrator / master | The parent agent that decomposes the task and delegates. Picks which specialist receives which sub-task. |
| **Leonardo** | Planner / strategist | Given an under-specified goal, produces a structured plan. Returns a list of steps with dependencies. |
| **Donatello** | Researcher / engineer | Operates RAG pipelines, web search, codebase inspection, document analysis. Returns evidence with citations. |
| **Michelangelo** | Writer / communicator | Produces user-facing artifacts: prose, code, slides, summaries. Adapts register to the audience. |
| **Raphael** | Executor | The agent with side-effect privileges: file writes, API calls, deployments, irreversible operations. Gated by the strictest allow-list. |

*Table 2: the Ninja Turtles multi-agent role taxonomy — one specialist Turtle per role, each shipped as its own CLI.*

A typical orchestration: Splinter receives the task, asks Leonardo for a plan, dispatches Donatello to gather evidence, hands the evidence and plan to Michelangelo to draft an artifact, and finally invokes Raphael to commit / deploy / send. Each step is a `call_cli` invocation from Splinter; each child is a Turtle CLI; each has its own session, its own scaffolding, its own allow-list. Concretely, from Splinter's perspective the delegation is nothing more than calling other CLIs:

```
# Splinter (orchestrator) — each delegate is just another CLI
plan=$(call_cli leonardo "plan: migrate auth to OAuth-over-MCP" --json)
evidence=$(call_cli donatello "gather: current auth code paths + the OAuth-MCP
spec" --json)
draft=$(call_cli michelangelo "write the migration guide" --plan "$plan" --
evidence "$evidence")
call_cli raphael "apply the migration" --plan "$plan" --depth 1   # side effects:
strictest allow-list
```

*Listing 4: Splinter orchestrating the specialist Turtles — delegation is plain `call_cli` composition, with the plan passed as an ordinary shell variable.*

The plan flows from Leonardo into Raphael as an ordinary shell variable; no shared typed-state object, no framework message bus. The naming homage is a cultural reference that also reinforces the through-line to Papert's Logo turtle via the Turtle pattern (§3.4).

### 5.3 Ralph loops as a sub-case

The convergence-loop pattern named after Geoffrey Huntley's *Ralph* essay [Huntley 2026] is trivially expressible as a CLI in this methodology, because it *is* bash:

```
# ralph.sh — convergence loop, ship as a CLI tool
goal="$1"; check_cli="$2"; max_iter="${3:-10}"
session=$(uuidgen)
for i in $(seq 1 "$max_iter"); do
  turtle "$session" "make progress on: $goal"
  "$check_cli" && { echo "converged in $i iter"; exit 0; }
done
echo "did not converge in $max_iter iter"; exit 1
```

*Listing 5: The Ralph loop, packaged as a CLI.* The orchestrator invokes `call_cli ralph "fix bug X" check_bug.sh 10`. The convergence check is the handoff protocol: the child writes the artifact; the check reads it; the loop continues based on the check's exit code. **Bash semantics carry the orchestration logic.** No framework primitive required.

### 5.4 Where the corollary leaks — what frameworks still do better

The corollary covers a substantial portion of multi-agent need, but it is not complete. Six places where a framework still offers something the bare corollary does not — in some a genuine missing capability, in others only turnkey convenience over equivalent engineering:

1. **Shared typed state across agents.** Frameworks like LangGraph [LangChain 2026b] give typed `StateGraph` abstractions where multiple agents share a structured object. CLI composition requires the engineer to build the equivalent with files on disk or a SQLite blackboard — solvable, but engineering effort rather than provided primitive.
2. **Streaming partial results across the orchestration tree.** A subprocess returns when it exits. For interactive user experience requiring per-token streaming from leaf to root, frameworks provide streaming primitives; CLI composition requires the engineer to rebuild the equivalent (typically with JSONL on stdout).
3. **Durable, suspendable, resumable workflows.** Workflow systems (Temporal [Temporal 2026], Inngest [Inngest 2026], restate.dev [Restate 2026]) provide state machines with retries, timeouts, idempotency, and durable suspend/resume. CLI orchestration provides none of those. For workflows with strict resume semantics, framework or workflow-engine territory.
4. **Typed handoff schemas.** Frameworks let the engineer declare a structured schema for inter-agent messages. CLI gives text or JSON; validation is on the caller.

5. **Unified observability across the tree.** Multi-agent tracing tools render a run as a unified graph; CLI composition gives independent per-agent transcripts that must be joined. This is the softest of the six, because the hook building-block (§3.2) closes it: a hook that emits a structured event on every tool call and child invocation into a shared append-only log lets ordinary tooling join the per-agent transcripts into one trace. The authors have built exactly such a layer for a multi-agent workshop — hooks emitting presence and advisory-lock events into a shared store, joined and rendered by a small script-and-skill toolchain, released as an open example [Legati Workshop 2026] — so for this item a framework buys *turnkey rendering*, not a capability the corollary otherwise lacks; it sits with (1) and (2) as engineering effort rather than a missing primitive.
6. **Recursion depth budgets.** A naïve sub-instance can spawn its own sub-instances. The methodology must declare a depth metric — either as an allow-list constraint or as a `--depth N` argument that decrements. The LLM will not set its own depth limit; the scaffolding has to. Concretely: the orchestrator passes `--depth N` on every child invocation; the child's pre-tool-use hook (§3.2) reads it, refuses any `call_cli` to another Turtle when `N ≤ 0`, and otherwise dispatches the grandchild with `--depth N-1`. The budget is enforced in host code on the way down, so a runaway delegation chain terminates at a fixed, auditable depth rather than at wherever the model happens to stop.

These are the criteria for **when to reach for a framework anyway** (see §7) — a genuine missing capability for the durable-workflow, typed-handoff, and depth-budget cases (3, 4, 6), and a convenience trade-off where the corollary is already reachable with modest engineering (1, 2, 5).

---

## 6. Internal deployment evidence — the AAC (Agent-Assisted Creator)

The AAC (Agent-Assisted Creator) is a custom agent for the LAMB educational platform [LAMB Project], built over roughly ten days (28 March – 6 April 2026) by a single developer working with an AI pair-programmer, and **in production at two universities since April 2026**: Universitat Politècnica de Catalunya (UPC, ~150 educator-creators) and the University of the Basque Country (UPV/EHU, ~50). It assists educators — not engineers — in creating, refining, and testing AI tutors within LAMB without leaving the platform. The build pre-dated the methodology's articulation; the methodology was crystallised by recognising the practice the AAC build had instantiated. The case-study evidence in this section is **internal deployment experience**; build logs, the issue tracker, and the production-deployment record are the primary sources. We summarise the mapping here and report results that are characterisable from internal logs rather than from externally reproducible benchmarks.

The mapping is exact enough to be diagnostic:

**P1 instantiation.** The four practical constraints (cost, hallucination, context, time) were live engineering pressures from the first prototype; cache-hygiene discipline was developed during the build as the AAC's token costs revealed where the cached prefix was being invalidated. The structural framing (`tools → system → messages`) crystallised post-hoc while the build was being structured into formal teaching material, when the build's empirical cost shape made the structural claim visible against Anthropic's prompt-caching documentation [Anthropic 2026b].

**P2 instantiation.** Function calling drives every AAC operation. The `lamb-cli` tool was developed in parallel with the agent as a liteshell, exposing the LAMB platform's assistant / model / knowledge-base / rubric operations. Skills, characters, and instruction sets were the artifacts that survived from the prototype's system prompt into production.

**P3 instantiation.** Claude Code drove the first AAC prototype against the live LAMB platform. The reconnaissance session revealed that no `lamb rubric` command existed despite assistants referencing rubrics — the first architectural gap, surfaced by the act of building.

**P4 instantiation.** The AAC ships as `aac`, a Turtle-pattern CLI. Sessions, replay (as trace, not bit-exact rerun), `--help` for the next agent — all in place. Security lives entirely in the scaffolding's allow-list. The architecture went through three crystallisations during P4, and the sequence is itself the clearest illustration of *architecture-discovered-by-building*:

1. **HTTP-only.** The agent called the LAMB application over loopback REST. This gave clean request/response boundaries, but it paid a network round-trip on every tool call and complicated local authentication.
2. **Direct-call.** The agent invoked the application's Python methods directly. This removed the round-trip, but it bypassed the application's own request pipeline — authentication, validation, middleware — which then had to be re-implemented agent-side and drifted out of sync.
3. **HTTP-with-ASGI.** The agent calls its own API through an in-process ASGI transport. This recovers the full request pipeline of (1) with the in-process speed of (2), and no open socket.

None of this was designed up front. Each architecture was adopted, exercised, and found wanting until the build surfaced the one that held.

**P5 instantiation.** Driving the AAC with a general-purpose agent (Claude Code) through realistic educator scenarios — exactly the agent-tests-agent pattern of §3.5 — surfaced a class of behavioural regressions that the narrow developer tests had missed, and that internal documentation logs as the "seven bugs." Each generalises to a P5 discipline and an anti-mandate. Tables 3 and 4 give four representative cases — Table 3 the failure and how P5 surfaced it, Table 4 the fix and the anti-mandate it generalises to; the full seven are recorded in the project's internal documentation.

| Bug | Failure (what the user hit) | How P5 surfaced it |
|---|---|---|
| **Confirmation loop** | User picks "1" to approve a write; the agent re-asks; the loop never terminates | The driving agent approving an assistant-creation through the numbered menu |
| **Jargon leak** | Agent tells an educator *"the debug didn't return useful pipeline output… prompt_processor is simple_augment…"* | The driving agent role-playing an educator asking why an assistant failed |
| **Internal fields in output** | Agent shows `Group ID: fed28308-…` (an internal integration id) to the educator | The driving agent requesting assistant details and inspecting the rendered output |
| **Session resume shows internals** | Reopening the page renders `[System: Skill startup]` as a user message | The driving agent resuming a stored session |

*Table 3: four representative AAC bugs and how P5 surfaced them — a general-purpose agent driving the custom agent through realistic educator scenarios.*

| Bug | Fix | Anti-mandate it teaches |
|---|---|---|
| **Confirmation loop** | Replace numbered options with explicit `(y)es / (n)o` | The LLM's output format and the host parser's expected input are one contract — do not let them drift |
| **Jargon leak** | An explicit *negative* vocabulary in the system prompt (name the forbidden words) | Positive instructions ("use simple language") are weaker than a named negative list — do not rely on them for vocabulary control |
| **Internal fields in output** | System-prompt rule listing which fields to hide vs. show | Do not pass tool results to the user unfiltered — the model's instinct is to display everything it receives |

| Bug | Fix | Anti-mandate it teaches |
|---|---|---|
| **Session resume shows internals** | Frontend filter: hide messages whose content starts with `[System:` | The stored (LLM) conversation and the displayed (user) conversation are different conversations — do not conflate them |

*Table 4: the fix and the anti-mandate each bug in Table 3 generalises to.*

The pattern across all seven is the methodology's central *encode-where-it-belongs* rule: critical constraints go in host code (deterministic), behavioural guidance goes in the system prompt, and a prompt rule that keeps being violated is promoted to a code rule. Tellingly, every one of the seven was a naming, format, vocabulary, or display defect at the human–system interface — not one was a crash or a timeout. That is precisely the behavioural surface classical testing does not exercise and P5 does.

The discipline did not stay with the build. In production, a principal use of the deployed AAC is exactly the testing move P5 names: educators use it to **automatically generate test sets for an assistant, then execute and evaluate them** — and at UPC nearly half of the ~150 creator-users now build tests this way. Agent-assisted behavioural testing is thus not only *how the AAC was built*; it is a feature that non-engineers use, at scale, on their own assistants. The methodology's testing posture has transferred from the engineer to the end user — which is the strongest evidence we can offer that the posture is usable rather than merely advocated.

The AAC's diagnostic value for the paper: it was built with the right discipline by intuition; the methodology makes the intuition explicit. A reader following the methodology arrives at the AAC's shape on purpose, not by accident.

### 6.1 Transfer beyond the AAC — agents building agents

The AAC is a single case built by the methodology's own originators, which is exactly the limitation a reviewer should press (and §7.5 concedes). The methodology's first test *outside* that team is now under way, and it happens to be the recursion the title names. A developer who did not build the AAC was trained in the methodology and applied it to a different LAMB subproject: a Moodle-CLI integration and a rubric-based evaluation agent, built on the published proof-of-concept for generative-AI-assisted rubric evaluation in software-engineering coursework [García-Peñalvo et al. 2026]. Following the methodology, that developer produced a CLI agent that does two things — it *applies* the methodology, and it *builds further agents*: evaluation agents that run complex extract → analyse → synthesise rubric assessments, generated for new domains. The agent-generated evaluators were validated against the original Python-plus-LLM pipeline on matched inputs, comparing final assessments.

This is *agents all the way down* in the literal, operational sense — an agent built under the methodology building the next layer of agents — and it is, equally, the methodology's **transferability evidence**: the practice was reproduced, and extended, by someone other than its authors, which is a different and stronger claim than "the authors applied their own method successfully." We report it as **in progress**: the work is pending publication and will be defended as an undergraduate final project (TFG) in late June 2026, with the report to follow. We therefore present it as forthcoming corroboration rather than a completed study, and flag it as such in §7.5. The same transfer pattern is being widened deliberately: a cohort on a Barcelona Supercomputing Center course (June–July 2026) will be the next, larger test of whether trained practitioners reproduce the results.

---

## 7. Discussion

### 7.1 Frameworks remain the right choice — sometimes

The methodology does not argue against frameworks. It argues against **adopting a framework before P1 and P2 are in the engineer's hands**. Once the substrate and building blocks are internalised, a framework is one of several engineering tools the practitioner can deploy with judgement.

A framework remains the right choice when:

- The problem formulation the framework encodes is genuinely the engineer's problem formulation. (LangGraph's typed StateGraph if orchestration genuinely needs typed shared state; AutoGen's conversation patterns if agents genuinely converse rather than delegate.)
- The team coordination cost of the unframed alternative exceeds the technical cost of the framework. A framework's value is sometimes located in the team layer rather than the technical layer — the cost of onboarding, code review, and developer rotation drops when the team standardises on a known framework.
- The cases identified in §5.4 — durable workflows, typed handoffs, streaming partials, recursion-depth control — where the framework writes plumbing the engineer would otherwise rebuild, and the engineering cost of reinventing it is unjustifiable. (Unified observability is the deliberate exception: §5.4 shows the hook building-block closes it cheaply, so a framework there buys turnkey rendering, not a missing capability.)

The methodology's claim is more modest than "frameworks are bad". It is that frameworks teach the framework, not the principles. A framework adopted competently after the principles are in hand is a different artifact than a framework adopted before. The methodology supports the first; it cautions against the second.

### 7.2 Minimal dependencies as a security posture

Recent supply-chain incidents make dependency risk a routine engineering concern rather than a purely hypothetical one. In 2024, the `xz`-utils backdoor (CVE-2024-3094) [NVD 2024] demonstrated how a core dependency could be compromised upstream — with adversarial maintainership concealed for an extended period before discovery. In 2026, PyPI documented package-supply-chain attacks affecting LiteLLM and Telnyx [PyPI 2026]. PyPI has also reported sustained malware volume and ongoing typosquatting defences [PyPI 2023]. These incidents do not prove that framework-heavy designs are always less secure, but they do strengthen the engineering case for understanding and minimising dependency closure where possible.

A custom agent built on the methodology proposed here typically depends on:

- One LLM-provider SDK (single package, audited by the provider).
- Whatever the application's existing dependencies are.
- No agent framework.

A custom agent built on a typical multi-framework agent stack (LLM orchestration framework + multi-agent framework + vector-database adapter + MCP client + evaluation framework) will typically pull in dozens of transitive dependencies. A direct empirical comparison is beyond this paper's scope; a controlled study across roughly a dozen matched agents — measuring transitive dependency count, lockfile size, and known-CVE exposure per stack — would be enough to settle it, and we flag it as future work (§7.4). The directional point is nonetheless sound: **each dependency introduces additional attack surface and maintenance obligations, so smaller dependency closures can reduce some classes of supply-chain risk**, independent of any judgement about the framework's correctness or its authors' trustworthiness.

We do not argue this as the primary reason to follow the methodology — the methodological reasons (substrate internalisation, agent-tests-agent as a discipline, the CLI orchestration corollary) are independent. We argue it as a *consequential* benefit: an agent built by the methodology proposed here has, by construction, a small dependency surface. In a 2026 threat environment where supply-chain attack is now a routine engineering assumption, this is a material design consideration.

### 7.3 Memory and the build / deploy asymmetry

The methodology presented in this paper covers building a custom agent. It does not cover what happens to that agent's memory once the build is done. The omission is a scope choice; this section names it explicitly so the reader sees where the methodology ends and where future work begins.

The methodology produces a class of artifact we call a **Turtle** — a custom agent whose memory exists only in the `messages[]` accumulated within a session, and whose tools, system prompt, skills, characters, and instruction set are frozen at deploy. A Turtle does not learn between sessions; whatever has been internalised by its design has been internalised in advance.

The agent that *builds* the Turtle is a different class of artifact. We adopt the role-name **Splinter** for it, following the same naming idiom as the Turtle pattern itself (§3.4) and the multi-agent topology of §5.2. A Splinter carries durable memory across sessions, evolves its form as the build proceeds, and has no fixed deploy moment. Concretely, a Splinter is whatever general-purpose orchestrator drives the engineer through P3 — Claude Code, OpenCode, or Cursor for an individual engineer; named long-running orchestrator agents for a team that maintains them across projects. *Different categories of artifact, not different gradations of the same thing.*

Across the P3 → P4 → P5 cycle, the Splinter stores, retrieves, and improves Turtle-relevant memories: what worked in the prototype, what scenarios surfaced behavioural defects in P5, what skill formulations the model misread, what voice and tone landed and what did not. At deploy, the Splinter hands the frozen state to the Turtle and steps out of the runtime. *The learning happened during the cycle and got compiled into the Turtle's deploy-time state.* The Turtle's runtime is then deliberately memoryless across sessions — which is what makes its behaviour governable, observable, and testable under the P5 discipline.

The recursion implied by the paper's title surfaces here in a second sense. *Agents all the way down* is also *memory-asymmetry all the way down*: at every level of agent composition, the agent doing the building carries durable memory; the agent being built carries session memory. The Splinter that helped build the Turtle is itself, in another setting, a Turtle that some prior Splinter — a developer guiding it, an orchestrator framework around it — once helped shape. The methodology's recursion includes this asymmetry as a structural feature, not a coincidence of any particular implementation.

This paper is therefore *the methodology for building a Turtle*. A companion direction — left to future work — is **the methodology for building a Splinter**: an agent whose memory is durable, whose form evolves across deployment, whose acceptance criteria, testing protocols, and security posture all differ from a Turtle's. That methodology is a separate paper, with separate engineering discipline. The contribution of this subsection to the present paper is the more modest one of naming the gap: a vocabulary item (*Turtle*, *Splinter*) and an explicit scope statement that says what this paper does and does not claim.

### 7.4 Open empirical questions

Three places where the methodology proposes claims that warrant systematic empirical work:

1. **Depth-2 orchestration cost shape.** The §5.4 cracks list claims a sub-instance of a sub-instance multiplies cold-cache prefixes. The systematic measurement — for which task sizes does fan-out pay off versus a single longer parent session — is open.
2. **P5 backbone-comparison protocols.** The methodology claims that running the same P5 scenario suite against different LLM backbones yields apples-to-apples behavioural comparison. The protocol — how to control for non-determinism, how to weight per-scenario outcomes, how to detect cross-backbone behavioural drift — is methodologically open.
3. **The minimal-dependencies / methodology-fidelity correlation.** The claim that agents built by the methodology are structurally simpler than framework equivalents is plausible; a controlled comparison (dependency count, lines of agent-loop code, cache-hit rate under iteration) on matched use cases would convert plausibility into evidence.

### 7.5 Threats to validity

We state the main limitations plainly. **Single primary case study.** The methodology was crystallised from one build, the AAC; although it has since transferred to further LAMB subprojects (§6), the bulk of the reported evidence comes from a single project and team, so generalisation beyond it is argued rather than measured. **Domain.** That project sits in EdTech; the substrate, building-block, and testing claims are domain-independent in principle, but we have not demonstrated them on agents in, for example, fintech or healthcare with their distinct compliance surfaces. **No controlled comparison.** We do not compare teams that follow the methodology against teams that do not, nor framework-free against framework-based builds on a matched task; the claims that the ordered phases and the iteration cycle yield better agents are reported as construction-period observations and practitioner judgement, not as controlled results. **Practitioner-sourced quantitative anchors.** The MCP-versus-CLI figures (§3.2) come from practitioner reports rather than peer-reviewed studies; we present them with their measurement conditions and ranges, and flag a controlled measurement as future work. These limitations set the agenda for the journal version of this work (§7.4).

---

## 8. Conclusion

We have proposed *Agents All the Way Down* — a methodology in five phases for building custom AI agents, from substrate internalisation through agent-tests-agent as a quality-assurance discipline. The methodology is lightweight (no required framework dependencies), comprehensive (covers substrate, building, prototyping, shipping, and testing), and independent (transferable across LLM providers, programming languages, and application domains). The corollary in §5 extends the methodology to multi-agent orchestration by recognising that the agent's CLI shipping artifact (P4) is itself the orchestration interface.

The methodology is the answer to a question the published literature has not framed cleanly: *what is the practice for building a custom agent, end to end, in 2026?* Existing publications give parts of the answer — substrate intuitions from Anthropic's engineering writing; tool-use mechanics from Anthropic's and OpenAI's documentation; prototyping affordances from Claude Code and OpenCode; orchestration patterns from LangGraph and AutoGen; evaluation patterns from Reflexion and the eval-tooling community. To our knowledge, no published practice has chained these into one.

The broader implication is for how software engineering treats agents at all. If a custom agent is a *living artifact* — its behavioural quality maintained by iteration against an evolving model substrate, never certified at a single point in time — then the disciplines that matter shift accordingly: from one-time acceptance toward continuous behavioural evaluation, from framework selection toward substrate fluency, from shipping a finished system toward maintaining a loop. P5 and the cycle of §4 are where that shift becomes concrete; the rest of the methodology exists to make the loop cheap enough to run in practice. This reframing — the agent as a maintained artifact rather than a delivered one — is, we think, the part most likely to outlast any specific tool named in this paper.

We close with two disciplines this paper hands the reader to try on Monday morning: (i) **ship the agent as a CLI** — the simplest single change with the largest downstream effect, because it makes everything else in this methodology composable; and (ii) **drive the agent with another agent**, in scenarios, against an evaluation rubric — the simplest single addition to a testing stack that classical software-engineering practice does not by itself provide. If those two disciplines are the only artifacts a reader takes from this paper, the reading was not wasted.

---

## Data availability

The methodology's primary worked example, the AAC (Agent-Assisted Creator), is part of the open-source **LAMB** platform (lamb-project.org); the public repository includes the AAC agent. Teaching material that walks the methodology through worked examples is openly available as a video lecture series, *Learning LLMs, RAG, and Building Agents* [Alier 2026]. The multi-agent observability layer discussed in §5.4 is released as an open example [Legati Workshop 2026].

---

## Acknowledgments

The methodology was developed across the following collaborations and communities. We thank:

- The wider **LAMB project** team at Universitat Politècnica de Catalunya (UPC) and Universidad del País Vasco / Euskal Herriko Unibertsitatea (UPV/EHU) (lamb-project.org) — the engineering ground that crystallised the methodology before it had a name, and the colleagues and students who have carried the AAC into production deployment at two universities since April 2026.
- **Raimon Lapuente** for the collaboration that helped structure the methodology into formal teaching material.
- The students and practitioners who engaged with early presentations of this material, whose questions and pushback sharpened the framing of P3–P4.
- The open agent community — practitioner writing by Geoffrey Huntley (Ralph-loop framing), Simon Willison (sustained practitioner-grade documentation of the LLM tooling landscape), and the Claude Code and OpenCode teams for normalising the general-purpose code agent.
- The Logo and constructionism tradition — Seymour Papert, the original Turtle, and the educational-computing lineage that gave the methodology's mascot its name.

---

## Appendix A — Acronyms and abbreviations

This paper draws on agent-engineering, software-engineering, and security vocabularies that overlap unevenly. The table below collects the acronyms used in the body in one place. Inline expansions appear at first use.

| Acronym | Expansion | Context in this paper |
|---|---|---|
| **AAC** | Agent-Assisted Creator | The custom agent that crystallised the methodology; the worked example of §6. |
| **AI** | artificial intelligence | The umbrella term; "AI agent" throughout. |
| **API** | application programming interface | Function-calling APIs, REST APIs, LLM-provider APIs. |
| **ASGI** | Asynchronous Server Gateway Interface | A Python web-server interface; AAC uses an in-process ASGI transport to call its own API (§6). |
| **arXiv** | (proper noun) | The preprint server; not an acronym. |
| **CI/CD** | continuous integration / continuous deployment | Standard software-delivery terminology used in the worked-example list of §1.2. |
| **CLI** | command-line interface | Either a packaged CLI binary (`gh`, `gws`) or a CLI surface that an agent can drive; the *Turtle pattern* (§3.4) ships agents as CLIs. |
| **CVE** | Common Vulnerabilities and Exposures | The vulnerability-identifier system; e.g. CVE-2024-3094 in `xz` (§7.2). |

| Acronym | Expansion | Context in this paper |
|---|---|---|
| **DSPy** | (a framework name, not expanded) | One of several agent-construction frameworks referenced in §2. |
| **GoF** | Gang of Four | Informal shorthand for Gamma, Helm, Johnson and Vlissides, authors of *Design Patterns* [Gamma et al. 1994]. |
| **HTTP** | Hypertext Transfer Protocol | Network protocol; "HTTP-based transports" appear in the MCP authorisation discussion (§3.2). |
| **IEEE** | Institute of Electrical and Electronics Engineers | Publisher of *IEEE Software*; the journal whose CFP is cited in §2.5. |
| **JSON** | JavaScript Object Notation | Data interchange format; agent tool calls return JSON-shaped results. |
| **JSONL** | JSON Lines | One JSON object per line; used as a streaming format for CLI inter-agent composition (§5.4). |
| **KV cache** | Key-Value cache | Internal cache of LLM keys-and-values for the prompt prefix; central to the §3.1 cost discussion. |
| **LAMB** | Learning Assistants Manager and Builder | The open-source educational AI platform the AAC was built for; in production at two universities since April 2026. |
| **LLM** | large language model | The underlying model the agent drives. |
| **MCP** | Model Context Protocol | Open standard for connecting LLM applications to external tools and context |

| Acronym | Expansion | Context in this paper |
|---|---|---|
| | | [Anthropic 2024b; MCP 2025]. |
| **NVD** | National Vulnerability Database | The U.S. government vulnerability registry; the CVE-2024-3094 advisory cited in §7.2 is housed there. |
| **OAuth** | Open Authorization | Standard delegated-authorisation framework; the MCP authorisation discussion (§3.2) cites OAuth-over-MCP transport behaviour. |
| **PyPI** | Python Package Index | The Python software repository; supply-chain incident records in §7.2. |
| **RAG** | Retrieval-Augmented Generation | A pattern in which an LLM is given retrieved-document context before answering; appears in the §5.2 multi-agent role taxonomy (Donatello operates RAG pipelines). |
| **REST** | Representational State Transfer | Web-service architectural style; "REST API" usage throughout. |
| **SDK** | software development kit | A provider-supplied library (e.g. the Anthropic SDK, the OpenAI SDK) used by the agent (§7.2). |
| **SE** | software engineering | As in "classical SE testing" (§3.5) or "agentic SE" (§2.5). |
| **STDIO** | standard input/output | A local-process transport for MCP; treated separately from HTTP trans- |

| Acronym | Expansion | Context in this paper |
|---|---|---|
| | | ports under MCP authorisation (§3.2). |
| **UPC** | Universitat Politècnica de Catalunya | Marc Alier and María José Casañ Guerrero's home institution. |
| **UPV/EHU** | Universidad del País Vasco / Euskal Herriko Unibertsitatea | Juanan Pereira's home institution (Donostia–San Sebastián campus). |
| **URL** | Uniform Resource Locator | Web address. |
| **USAL** | Universidad de Salamanca | Francisco José García-Peñalvo's home institution. |

**Methodology-specific terms** introduced in this paper (defined in the body where they first appear, listed here for reference):

| Term | Where defined | One-line gloss |
|---|---|---|
| **Liteshell** | §3.2 | An in-process facade that presents an application's operations to the LLM as if they were a CLI; carries the §3.2 CLI-token-efficiency win into cloud-native deployments. |
| **Hook** | §3.2, §3.4 | A deterministic host-side function that fires on a fixed point in the agent's lifecycle (notably `pre-tool-use`); the enforcement mechanism for the scaffolding's security check — where the allow-list is actually applied (Listing 2). |
| **Turtle pattern** | §3.4 | The P4 deployment shape: the custom agent ships as a CLI tool with stable invocation, session resume, structured outputs, and `--help`. |

| Term | Where defined | One-line gloss |
|---|---|---|
| **Turtle corollary** | §5 | Multi-agent orchestration as CLI composition, derived from the Turtle pattern. |
| **Agent-tests-agent** | §3.5 | The P5 discipline: a general-purpose agent drives the custom agent through scenarios; behavioural-evaluation layer above classical SE testing. |
| **Substrate / Building blocks / Prototype / Harvest, fold, ship as CLI / Agent-tests-agent** | §3 | The five phases. P1–P2 are preconditions; P3–P5 are the iterated practice (§4). |
| **Cabinet maker's jig vs. carpenter's bench saw** | §1.1 | The fit-not-capability distinction between a custom agent and the general-purpose tier. |

---

## References

The references below are in author-year format suitable for markdown reading; the BibTeX equivalents are listed at the end of this section for the eventual arXiv / TeX submission.